\documentclass{article}

\usepackage{a4wide, amsmath,amsthm,amsfonts,amscd,amssymb,eucal,bbm,mathrsfs}
\usepackage{hyperref}

\def\RR{{\mathbb R}}

\def\A{{\mathcal A}}

\def\1{{\mathbbm 1}}

\def\u1{{\A^{(0)}}}

\def\diff{{\rm Diff}}

\def\diffs1{\diff(S^1)}
\def\mob{{\rm M\ddot{o}b}}
\def\mob2{{\rm M\ddot{o}b}^{(2)}}

\def\psl2r{{\rm PSL}(2,\RR)}
\def\2dmob{{\overline{\psl2r}\times\overline{\psl2r}}}
\def\<{\langle}
\def\>{\rangle}

\newtheorem{theorem}{Theorem}[section]

\theoremstyle{remark}

\title{Construction of wedge-local QFT through Longo-Witten endomorphisms\footnote{This article
has been prepared for the proceedings of ICMP 12.}}
\date{}
\author{
{\bf Yoh Tanimoto}\footnote{Supported by Deutscher Akademischer Austauschdienst
and in part by Courant Research Centre ``Higher Order Structures in Mathematics''.} \\
e-mail: {\tt yoh.tanimoto@theorie.physik.uni-goettingen.de}\\
Institut f\"ur Theoretische Physik, Universt\"at G\"ottingen,\\
Friedrich-Hund-Platz 1, 37077 G\"ottingen, Germany.\\
}

\begin{document}
\maketitle

\begin{abstract}
We review our recent construction of operator-algebraic quantum field models
with a weak localization property.
Chiral components of two-dimensional conformal fields and certain endomorphisms
of their observable algebras play a crucial role. In one case, this construction
leads to a family of strictly local (Haag-Kastler) nets.
\end{abstract}

\section{Operator-algebraic approach to QFT}\label{oa}
The problem of constructing interacting relativistic quantum field theory on four-dimensional
spacetime has been a long-standing open problem. On the other hand, in lower dimensions
there have been important developments in several different approaches (e.g.\! Constructive QFT,
form factor bootstrap).
Here we adopt the operator-algebraic approach, also known as Algebraic Quantum Field Theory or
Local Quantum Physics \cite{Haag96}, which has recently resulted in a construction of a large
family of quantum field models in two-spacetime dimensions \cite{Lechner08} and a further
progress \cite{Lechner10}. Furthermore, two-dimensional Conformal Field Theory can be
successfully studied in this framework \cite{Kawahigashi05}. In this contribution, we present
a new method of constructing two-dimensional operator-algebraic QFT based on chiral components of
CFT \cite{Tanimoto12-2, BT12, Tanimoto12-3}.

The mathematical approach to QFT which is the closest to the notion in physics is the Wightman
axioms. A Wightman field is an operator-valued distribution which satisfies certain properties
which originate in physics, e.g.\! Poincar\'e covariance and Einstein causality. Yet, the quantum
field smeared with a smooth function is still an unbounded operator and sometimes plagued by the
problem of domains. Instead, in algebraic QFT one considers algebras of bounded operators.

\subsection{Haag-Kastler nets}
A {\bf Haag-Kastler net}, or a {\bf Poincar\'e covariant net (of observables)} assigns to each
open region $O \subset \mathbb{R}^d$ a von Neumann algebra $\mathcal{A}(O)$ on a common Hilbert space $\mathcal{H}$
(see the book \cite{TakesakiI} for a general account on von Neumann algebras).
In addition, one assumes that there is a continuous unitary representation $U$
of the Poincar\'e group on $\mathcal{H}$ and an invariant ground state, the vacuum $\Omega$.
The triple $(\mathcal{A},U,\Omega)$ is subject to standard axioms and considered as
a model of quantum field theory \cite{Haag96}.

If one has a Wightman field $\phi$, then one can construct the corresponding net by
defining ${\mathcal A}(O) := \{e^{i\phi(f)}: {\rm supp} f \subset O\}''$, where ${\mathcal M}'$ means the set
of bounded operators commuting with any element of ${\mathcal M}$. The double commutant ${\mathcal M}''$
is the smallest von Neumann algebra which includes ${\mathcal M}$. Actually it is required that
$\phi(f)$ and $\phi(g)$ have commuting spectral projections for $f, g$ with spacelike
separated support, but in many cases this is satisfied.

Conversely, when a Haag-Kastler net ${\mathcal A}$ is given and if certain technical conditions
are satisfied, then one can recover quantum fields. In this way, a Haag-Kastler net is considered
as the operator-algebraic formulation of quantum field theory.

Now we are concerned with the construction problem. In any spacetime dimension, there are
the so-called (generalized) free fields, and corresponding nets. But other examples are
rare. For $d\le 3$ the methods of Constructive QFT have successfully constructed
many examples or for $d=2$ there are plenty of examples of conformal fields,
but apart from these models, it is very hard to construct such fields or nets.
In this work we address this problem for $d=2$ with a purely von Neumann algebraic approach.
An extension to higher dimensions remains open.

\subsection{Borchers triples}
One of the difficulties in constructing Haag-Kastler nets lies in the infiniteness of the family
$\{{\mathcal A}(O)\}$ with certain compatibility conditions. Instead, Borchers observed that for $d=2$,
actually the whole net can be recovered from the single von Neumann algebra ${\mathcal A}(W_{\mathrm R})$
associated with the (right-)wedge-shaped regions $W_{\mathrm R} := \{a \in {\mathbb R}^2: a_1 > |a_0|\}$ and the
spacetime symmetry $U$ (under the condition called Haag-duality). Furthermore, by the
Tomita-Takesaki theory of von Neumann algebras \cite{TakesakiII}, it is enough to know
the restriction of $U$ to the translation subgroup ${\mathbb R}^2$ \cite{Borchers92}.

A {\bf Borchers triple} $({\mathcal M},T,\Omega)$ consists of a von Neumann algebra ${\mathcal M}$ on ${\mathcal H}$, a unitary
representation $T$ of ${\mathbb R}^2$ with joint spectrum in $V_+$ and a vacuum vector $\Omega$
such that $\Omega$ is invariant under $T(a)$, ${{\rm Ad\,}} T(a) {\mathcal M} \subset {\mathcal M}$ for $a \in W_{\mathrm R}$
and ${\mathcal M}\Omega$ and ${\mathcal M}'\Omega$ are dense in ${\mathcal H}$ (these properties are called
{\bf cyclicity} and {\bf separating property} of $\Omega$ for ${\mathcal M}$, respectively).
It is easy to see that if $({\mathcal A},U,\Omega)$ is a Poincar\'e covariant net,
then $({\mathcal A}(W_{\mathrm R}), U|_{{\mathbb R}^2}, \Omega)$ is a Borchers triple.

Conversely, starting with a Borchers triple $({\mathcal M},T,\Omega)$, one can define a net
as follows: in two-spacetime dimensions, any double cone can be represented as the intersection
of two-wedges $(W_{\mathrm R}+a)\cap (W_{\mathrm L}+b) =: D_{a,b}$, where $W_{\mathrm L}$ is the reflected (left-)wedge.
Then one defines first von Neumann algebras ${\mathcal A}(D_{a,b})$ for double cones $D_{a,b}$
by ${\mathcal A}(D_{a,b}) := {{\rm Ad\,}} T(a)({\mathcal M}) \cap {{\rm Ad\,}} T(b)({\mathcal M}')$. For a general region $O$ one takes
${\mathcal A}(O) := \left(\bigcup_{D_{a,b}\subset O} {\mathcal A}(D_{a,b})\right)''$. Then one can show that
this ``net'' ${\mathcal A}$ satisfies isotony and locality. Furthermore, the representation $T$
extends to a representation $U$ of the Poincar\'e group which makes ${\mathcal A}$ covariant and
$\Omega$ is still invariant. In this way one obtains a ``net'' $({\mathcal A}, U, \Omega)$,
where the only missing property is that $\Omega$ is cyclic for ${\mathcal A}(O)$.
In general, if one starts with a net, goes down to the Borchers triple and back to
the net as above, this does not coincide with the original net, but such a difference
is not important when one is interested in the construction of an interacting net.

Hence, in the operator-algebraic approach, the construction of Haag-Kastler nets can be
split into two steps: (1) to construct Borchers triples, (2) to prove the cyclicity of $\Omega$.
In the following, we carry out (1) for massless models in Section \ref{massless} and
both (1) and (2) for massive models in Section \ref{massive}.

\section{Conformal nets and Longo-Witten endomorphisms}\label{conformal}
Here we review the operator-algebraic treatment of Conformal Field Theory, which
is the main ingredient of our construction of Borchers triples. As is well known,
two-dimensional CFT contains observables which are invariant under right- or
left-lightlike translations. They are called chiral components and can be considered
as observables defined on lightrays. They can often be extended to $S^1$, the one-point
compactification of a lightray.
A {\bf M\"obius covariant net} $(\mathcal{A}_0, U_0,\Omega_0)$ on $S^1$ is
defined precisely like a Haag-Kastler net in $\mathbb{R}^d$,
with the regions taken to be intervals in $S^1$, and the Poincar\'e group is replaced by
the M\"obius group $\mathrm{PSL}(2,\mathbb{R})$.

There are many examples of M\"obius covariant nets: the $U(1)$-current net (free boson), the free
fermion net, the Virasoro nets, Minimal models, WZW models, etc. However, one cannot define
the notion of interaction for one-dimensional nets (actually, it can be shown that
a two-dimensional CFT does not interact in the sense of scattering theory \cite{Tanimoto12-1}).
Rather, they are building blocks of interacting two-dimensional QFT. Most important ones are
the $U(1)$-current and the free fermion, which admit Fock space structure.
The aim of the sequel is to demonstrate new methods to construct Borchers triples and Haag-Kastler nets
on $\mathbb{R}^2$ from M\"obius covariant nets on $S^1$.

Let us now introduce the main notion in our construction. Recall that the circle $S^1$ is
identified with ${\mathbb R}\cup\{\infty\}$ by the stereographic projection.
A {\bf Longo-Witten endomorphism} of a conformal net is an endomorphism of ${\mathcal A}_0({\mathbb R}_+)$,
implemented by a unitary $V_0$ which commutes with translation $T_0$.
This notion was first introduced in order to construct QFT with boundary \cite{LW11}.

The simplest example of Longo-Witten endomorphisms is the translation itself ${{\rm Ad\,}} T_0(t)$,
$t\ge 0$. In many examples there are {\bf inner symmetries}, for which there is a unitary
$V_0$ such that ${{\rm Ad\,}} V_0({\mathcal A}_0(I)) = {\mathcal A}_0(I)$ and $V_0\Omega_0 = \Omega_0$. Then
$V_0$ automatically commutes with $U_0(g)$, especially with $T_0(t)$ and implements
a Longo-Witten endomorphism.

The examples above are rather of general nature. By considering a specific net, the
$U(1)$-current net ${\mathcal A}_{U(1)}$, Longo and Witten found a large family of examples \cite{LW11}.
The $U(1)$-current net is defined on the symmetric Fock space ${\mathcal H}_0$, on whose one-particle space
the M\"obius group ${\rm PSL}(2,{\mathbb R})$ acts irreducibly with lowest weight $1$. There is a operator-valued
distribution $J$, the current, which generates the net as explained above. One can promote
a unitary operator $V_1$ on the one-particle space ${\mathcal H}_1$
to a unitary operator $\Gamma(V_1)$ on the full space ${\mathcal H}_0$ ({\bf second quantization}).
The M\"obius group representation promotes to ${\mathcal H}_0$ as well.

An {\bf inner symmetric function} $\varphi$ is the boundary value of a bounded analytic function
on the upper-half plane, with $|\varphi(t)| = 1$ for $t\in {\mathbb R}$, and $\varphi(-t) = \overline{\varphi(t)}$.

\begin{theorem}[Longo-Witten]
 Let $\varphi$ be an inner symmetric function and $P_1$ be the generator of the one-particle
translation of the $U(1)$-current net ${\mathcal A}_{U(1)}$. Then the unitary operator $\Gamma(\varphi(P_1))$ implements
a Longo-Witten endomorphism of ${\mathcal A}_{U(1)}$.
\end{theorem}

Using these operators, we construct two-dimensional models in the next Sections.

\section{Massless construction and scattering theory}\label{massless}
\subsection{Massless scattering theory}
First we consider two-dimensional massless models. The reason is, as we see below,
the scattering theory is particularly simple and a general structural result can be
obtained. The following is the adaptation of the theory \cite{Buchholz75} to Borchers triples
\cite{DT11}.

Let $({\mathcal M}, T, \Omega)$ be a (two-dimensional) Borchers triple.
As explained before, an element $x \in {\mathcal M}$ should be considered as an observable
in the wedge region $W_{\mathrm R}$. If one defines the following,
\[
 x_\pm(h_{\mathcal T}) := \int h_{\mathcal T}(t) {{\rm Ad\,}} T(t,\pm t)(x)dt,
\]
where $h$ is a nonnegative smooth function with compact support with $\int h(t)dt = 1$,
${\mathcal T}$ is a nonzero real number and $h_{\mathcal T}(t) = |{\mathcal T}|^{-\epsilon}h(|{\mathcal T}|^{-\epsilon}(t-{\mathcal T}))$ with some fixed
number $0 < \epsilon < 1$. Then the strong limits $\Phi^{\rm out}_+(x) = \lim_{{\mathcal T}\to \infty} x_+(h_{\mathcal T})$ and
$\Phi^{\rm in}_-(x) = \lim_{{\mathcal T}\to -\infty} x_-(h_{\mathcal T})$ exist.
The operators $\Phi^{\rm out}_+(x)$ and $\Phi^{\rm in}_-(x)$ are called {\bf asymptotic fields}.
Similarly one can define $\Phi^{\rm out}_-(x')$ and $\Phi^{\rm in}_+(x')$ using an element $x' \in {\mathcal M}'$.
They have nice properties, so that $\Phi^{\rm out}_\pm(\cdot)$ and $\Phi^{\rm in}_\pm(\cdot)$ can be considered as operators
at far-future and far-past, respectively. In particular, $\Phi^{\rm out}_+(x)$ and $\Phi^{\rm out}_-(x')$
commute and act like operators in tensor product. More precisely, let ${\mathcal H}_\pm$ be
the space of vectors invariant under $T(t,\pm t)$. $\Phi^{\rm out}_\pm(\cdot)$ can be naturally
restricted to ${\mathcal H}_\pm$ respectively.
Vectors $\Phi^{\rm out}_\pm(\cdot)\Omega$ describe massless excitations
going out in $\pm$ directions, respectively and a natural tensor product structure can be given
to the subspace generated by such operators.

If $\{\Phi^{\rm out}_+(x)\Phi^{\rm out}_-(x')\Omega: x\in{\mathcal M}, x'\in{\mathcal M}'\}$ and
$\{\Phi^{\rm in}_+(x')\Phi^{\rm in}_-(x)\Omega: x\in{\mathcal M}, x'\in{\mathcal M}'\}$ are total in ${\mathcal H}$,
then we say that the Borchers triple $({\mathcal M},T,\Omega)$ is {\bf asymptotically complete}, or
in other words, the any vector in ${\mathcal H}$ can be interpreted as in- and out-scattering states.
For $\xi = \Phi^{\rm out}_+(x)$ and $\eta = \Phi^{\rm out}_-(x')$, let us write
$\xi\overset{\rm out}\times\eta = \Phi^{\rm out}_+(x)\Phi^{\rm out}_-(x')\Omega$.
The operation $\overset{\rm out}\times$ naturally extends to ${\mathcal H}_\pm$. Similarly we define
$\overset{\rm in}\times$.
Then the {\bf S-matrix} defined by
$ S\xi\overset{\rm out}\times\eta = \xi\overset{\rm in}\times\eta$
is a unitary operator. Note that, because of the nondispersive nature of two-dimensional
massless excitations, it is enough to consider only such products of two operators.

Let us assume that an asymptotically complete Borchers triple $({\mathcal M},T,\Omega)$ comes from a net $({\mathcal A},T,\Omega)$.
It is interesting to observe that the information of the S-matrix and asymptotic fields
is enough to recover the given ${\mathcal M}$. In fact, we have the following simple formula \cite{Tanimoto12-2}.
\begin{theorem} Under standard assumptions (and asymptotic completeness), it holds that
 ${\mathcal M} = \{\Phi^{\rm out}_+(x), {{\rm Ad\,}} S(\Phi^{\rm out}_-(x')): x\in{\mathcal M}, x'\in{\mathcal M}'\}''$.
\end{theorem}

As remarked before, $\Phi^{\rm out}_\pm(\cdot)$ are observables acting like in tensor product.
On the other hand, ${\mathcal M}$ is the wedge-algebra of the interacting theory. And the net can
be recovered from the Borchers triple, namely from ${\mathcal M}$ and translation.
Furthermore, it can be shown that $\Phi^{\rm out}_\pm(\cdot)$ generate one-dimensional
conformal net \cite{Tanimoto12-1, Tanimoto12-2}.
Hence the above formula tells us how to construct interacting theory out of free, conformal theory
and S-matrix. In the next Subsection we carry out this program.

\subsection{Construction of massless Borchers triples}
\subsubsection{One-parameter Longo-Witten endomorphisms}
Let us take a M\"obius covariant net $({\mathcal A}_0, U_0, \Omega_0)$ on ${\mathcal H}_0$. We denote by $T_0$ the
restriction of $U_0$ to the translation subgroup as before. With its positive generator $P_0$,
one can write $T_0(t) = e^{itP_0}$. Only with this setting, we can already construct nontrivial
Borchers triples \cite{Tanimoto12-2}.

The full Hilbert space is now the tensor product ${\mathcal H} := {\mathcal H}_0\otimes {\mathcal H}_0$.
For a point $(t_+,t_-)$ in the lightray coordinate of the Minkowski space,
we define $T(t_+,t_-) := T_0(t_+)\otimes T_0(t_-)$. Again the simple tensor product
$\Omega := \Omega_0\otimes \Omega_0$ plays the role of the vacuum vector.
\begin{theorem}
 Let $\kappa \ge 0$ and
\[
 {\mathcal M}_\kappa := \{x\otimes {\mathbbm 1}, {{\rm Ad\,}} e^{i\kappa P_0\otimes P_0}({\mathbbm 1}\otimes y): x\in{\mathcal A}_0({\mathbb R}_-), y\in{\mathcal A}_0({\mathbb R}_+)\}''.
\]
Then $({\mathcal M}_\kappa, T,\Omega)$ is an asymptotically complete Borchers triple with the S-matrix $e^{i\kappa P_0\otimes P_0}$.
\end{theorem}

The proof goes as follows. The conditions on $T$ and $\Omega$ are easily verified and
it is also simple to check ${{\rm Ad\,}} T(a) ({\mathcal M}_\kappa) \subset {\mathcal M}_\kappa$ for $a\in W_{\mathrm R}$, since
$T(a)$ and $e^{i\kappa P_0\otimes P_0}$ commute. The cyclicity of $\Omega$ for ${\mathcal M}_\kappa$ follows
immediately from the definition of conformal nets. The only nontrivial part is the
separation by $\Omega$. In order to show this, it is enough to find another von Neumann algebra
which commutes with ${\mathcal M}$ and for which $\Omega$ is cyclic. Such a von Neumann algebra is given by
\[
 {\mathcal M}_\kappa^1 := \{{{\rm Ad\,}} e^{i\kappa P_0\otimes P_0}(x'\otimes {\mathbbm 1}), {\mathbbm 1}\otimes y': x'\in{\mathcal A}_0({\mathbb R}_+), y'\in{\mathcal A}_0({\mathbb R}_-)\}''.
\]
Let us see that, as an example, $x\otimes {\mathbbm 1}$ and ${{\rm Ad\,}} e^{i\kappa P_0\otimes P_0}(x'\otimes {\mathbbm 1})$ commute.
Since one has $e^{i\kappa P_0\otimes P_0} = \int e^{it\kappa P_0}\otimes dE_0(t)$, where $E_0$ is the spectral
measure of $P_0$, and $e^{it\kappa P_0}$
{\it implements a Longo-Witten endomorphism}, one has
${{\rm Ad\,}} e^{i\kappa P_0\otimes P_0}(x'\otimes {\mathbbm 1}) = \int {{\rm Ad\,}} e^{it\kappa P_0\otimes P_0}(x')\otimes dE_0(t)$
and ${{\rm Ad\,}} e^{it\kappa P_0\otimes P_0}(x')$ commutes with $x$. In this way, Longo-Witten endomorphisms
enter the present construction.

Note that when $\kappa = 0$ then ${\mathcal M}_\kappa = {\mathcal A}_0({\mathbb R}_-)\otimes {\mathcal A}_0({\mathbb R}_+)$ and $S_\kappa = {\mathbbm 1}$, namely,
the simple tensor product results in a noninteracting triple.
It has been revealed that $({\mathcal M}_\kappa, T,\Omega)$ is equivalent to the BLS deformation \cite{BLS11}
of $({\mathcal M}_0,T,\Omega)$.

It is also possible to perform a similar construction with one-parameter inner symmetries \cite{Tanimoto12-2}.
It is remarkable that in this case, with additional technical conditions,
one can determine the intersection of
wedges. However, the intersection turns out to be trivial in a certain sense.

\subsubsection{U(1)-current as building blocks}\label{u1}

The Longo-Witten endomorphisms on the $U(1)$-current net explained above can be used to construct
Borchers triples as well. Yet the spirit is always the same: one has only to find an S-matrix
to twist the right component.

As recalled before, the Hilbert space ${\mathcal H}_0$ for the $U(1)$-current is the symmetric Fock space.
One considers first the unsymmetrized Fock space ${\mathcal H}^\Sigma = \bigoplus_m {\mathcal H}_1^{\otimes m}$.

We fix an inner symmetric function $\varphi$.
On ${\mathcal H}_1^{\otimes m}$, there act $m$ commuting operators
\[
\{{\mathbbm 1}\otimes\cdots\otimes \underset{i\mbox{-th}}{P_1}\otimes\cdots \otimes {\mathbbm 1}: 1\le i \le m\}.
\]
Recall that functional calculus has been used in the construction by Longo and Witten. Here we put:
\begin{itemize}
\item $P_{i,j}^{m,n} := ({\mathbbm 1}\otimes\cdots\otimes \underset{i\mbox{-th}}{P_1}\otimes\cdots \otimes {\mathbbm 1})
\otimes ({\mathbbm 1}\otimes\cdots\otimes \underset{j\mbox{-th}}{P_1}\otimes\cdots \otimes {\mathbbm 1})$,
which acts on ${\mathcal H}_1^{\otimes m}\otimes{\mathcal H}_1^{\otimes n}$, $1 \le i \le m$ and $1 \le j \le n$.
\item $S^{m,n}_\varphi := \prod_{i,j}\varphi(P_{i,j}^{m,n})$, where
$\varphi(P_{i,j}^{m,n})$ is defined by the functional calculus on ${\mathcal H}_1^{\otimes m}\otimes{\mathcal H}_1^{\otimes n}$.
\item $S_\varphi := \bigoplus_{m,n} S_\varphi^{m,n} = \bigoplus_{m,n} \prod_{i,j} \varphi(P^{m,n}_{i,j})$
\end{itemize}
Then one can show that $S_\varphi$ restricts to the symmetric Fock space and
can be decomposed into a direct integral of Longo-Witten unitaries.
As in the case of one-parameter endomorphisms, we can show the following.
\begin{theorem}
Let us put
\[
 {\mathcal M}_\varphi := \{x\otimes {\mathbbm 1}, {{\rm Ad\,}} S_\varphi({\mathbbm 1}\otimes y): x\in{\mathcal A}_{U(1)}({\mathbb R}_-), y\in{\mathcal A}_{U(1)}({\mathbb R}_+)\}''.
\]
Then $({\mathcal M}_\varphi, T,\Omega)$ is an asymptotically complete Borchers triple with the S-matrix $S_\varphi$.
\end{theorem}
Recently we found that this construction is equivalent to the deformation of massless free field
by Lechner \cite{Lechner11}. This will be presented elsewhere \cite{LST12}.

Note that the S-matrix of all these constructions preserve the particle number in the sense of
Fock space. It is also possible to construct examples which do violate this structure as follows
\cite{BT12}.

It is known that the $U(1)$-current net is identified as a subtheory of the free complex fermion net ${\mathcal F}$
(the boson-fermion correspondence). The free fermion net ${\mathcal F}$ is defined on the fermionic Fock space.
Similarly as above, by choosing an inner (this time not necessarily symmetric) function $\varphi$,
one can construct a Borchers triple with S-matrix $S_{\varphi,{\mathcal F}}$ with the ``free'' part ${\mathcal F}\otimes{\mathcal F}$.
Then $S_{\varphi,{\mathcal F}}$ can be restricted to the bosonic part ${\mathcal A}_{U(0)}\otimes {\mathcal A}_{U(0)}$.
Let us denote the restriction by $S_{\varphi,{\rm r}}$
\begin{theorem}
Let us put
\[
 {\mathcal M}_{\varphi,{\rm r}} := \{x\otimes {\mathbbm 1}, {{\rm Ad\,}} S_{\varphi,{\rm r}}({\mathbbm 1}\otimes y): x\in{\mathcal A}_{U(1)}({\mathbb R}_-), y\in{\mathcal A}_{U(1)}({\mathbb R}_+)\}''.
\]
Then $({\mathcal M}_{\varphi, {\rm r}}, T,\Omega)$ is an asymptotically complete Borchers triple with the S-matrix $S_{\varphi,{\rm r}}$.
The S-matrix $S_{\varphi,{\rm r}}$ does not preserve $n$-particle space of the bosonic Fock space.
\end{theorem}
So the S-matrix appears to represent particle production. However, the the cyclicity of the vacuum
for the algebras of bounded regions is not known. Furthermore, the concept of particle is to be
replaced by waves in massless two-dimensional models \cite{Buchholz75}.
Thus it is desired to construct massive models, where these issues can be settled.

\section{Massive construction and strict locality}\label{massive}
It is also possible to construct massive models using Longo-Witten endomorphisms.
We only briefly sketch the construction \cite{Tanimoto12-3}.

Let $(\mathcal{A}_{\mathrm c}, U_{\mathrm c}, \Omega_{\mathrm c})$ be the free
{\em massive complex} field net.
There is an action of $U(1)$ by inner symmetry.
One can take the generator $Q_{\mathrm c}$, such that
$V_{\mathrm c}(\kappa) = e^{i2\pi\kappa Q_{\mathrm c}}, \kappa \in \mathbb{R}$.
We consider the wedge algebra $\mathcal{M}_{\mathrm c} = \mathcal{A}_{\mathrm{c}}(W_{\mathrm{R}})$ and
the restriction $T_{\mathrm c}$ of $U_{\mathrm c}$ to the subgroup of translations.

Our new massive models are constructed on the tensor product Hilbert space
$\widetilde{\mathcal{H}}_{\mathrm c} := \mathcal{H}_{\mathrm c}\otimes \mathcal{H}_{\mathrm c}$.
There is a natural representation $\widetilde{T}_{\mathrm c} := T_{\mathrm c}\otimes T_{\mathrm c}$
and a vector $\widetilde{\Omega}_{\mathrm c} := \Omega_{\mathrm c}\otimes\Omega_{\mathrm c}$.
For $\kappa \in \mathbb{R}$, we define
$\widetilde{V}_{{\mathrm c},\kappa} := e^{i2\pi \kappa Q_{\mathrm c}\otimes Q_{\mathrm c}}$.

\begin{theorem}
Let us put
\[
 \widetilde{\mathcal M}_{{\mathrm c}, \kappa} := \{x\otimes {\mathbbm 1}, {{\rm Ad\,}} \widetilde{V}_{{\mathrm c},\kappa}({\mathbbm 1}\otimes y): x, y\in{\mathcal M}_{\mathrm c}\}''.
\]
Then $(\widetilde{\mathcal M}_{{\mathrm c},\kappa}, \widetilde T_{\mathrm c},\widetilde\Omega_{\mathrm c})$ is a Borchers triple.
\end{theorem}

We can show that this triple is indeed strictly local without using modular nuclearity \cite{BL04},
by directly proving the wedge-split property \cite{DL84, Lechner08},
hence one can further construct nontrivial a Haag-Kastler net $\widetilde{\mathcal{A}}_{{\mathrm c},\kappa}$.
Since $\widetilde T_{\mathrm c}$ is a massive representation, the scattering theory works well.
One can show that the S-matrix is nontrivial.
The S-matrix is factorizing and does not depend on the rapidity.

It should be noted that this procedure to construct nontrivial Haag-Kastler nets can be applied
to any net with either modular nuclearity or wedge-split property which admits an action of $U(1)$ 
by inner symmetry.
One can take the tensor product of one of the models by Lechner \cite{Lechner08}, instead of
the complex free field. One can even repeat the procedure by taking the constructed triple
$(\widetilde{\mathcal M}_{{\mathrm c},\kappa}, \widetilde T_{\mathrm c},\widetilde\Omega_{\mathrm c})$
as an input to construct further new nets.

We indicate another construction of Borchers triples without investigating strict locality \cite{Tanimoto12-3}.
Here the most important observation is the following: one considers the massive real free
field (net) and takes the right-wedge algebra $\mathcal{A}(W_{\mathrm{R}})$ and the restriction of $U$
to the positive lightlike translations, Then it is equivalent to the $U(1)$-current net.
The negative lightlike translations (in the past direction)
are redefined as a suitable one-parameter semigroup of
Longo-Witten endomorphisms $\mathrm{Ad\,} V_0(s)$
with negative generator (this will be investigated in more detail \cite{BT13}).
Furthermore one has a Longo-Witten endomorphism as in Section \ref{u1} for an inner
symmetric function $\varphi$.
By using $\varphi$, instead of a group action of $U(1)$, one can twist the tensor product
of the real free field net. In this way one can construct a family of Borchers triples
with two-particle S-matrix which depends on rapidity.

Models with particle production have not yet been obtained with this method.

\bibliographystyle{plain}

\end{document}